\documentclass[twocolumn]{aastex62}
\usepackage{xspace}

\usepackage{xspace}
\usepackage[caption=false]{subfig}
\usepackage{booktabs}

\newcommand{\Ha}{H$\alpha$\xspace}
\newcommand{\Hb}{H$\beta$\xspace}
\newcommand{\HII}{H\,{\sc ii}]\xspace}
\newcommand{\Hgamma}{H$\gamma$\xspace}
\newcommand{\Hdelta}{H$\delta$\xspace}
\newcommand{\OII}{[O\,{\sc ii}]\xspace}
\newcommand{\CIII}{C\,{\sc iii}]\xspace}
\newcommand{\CII}{C\,{\sc ii}]\xspace}
\newcommand{\CIV}{C\,{\sc iv}\xspace}
\newcommand{\MgII}{Mg\,{\sc ii}\xspace}
\newcommand{\OIII}{[O\,{\sc iii}]\xspace}
\newcommand{\NII}{[N\,{\sc ii}]\xspace}

\newcommand{\NeIII}{[Ne\,{\sc iii}]\xspace}

\newcommand{\NeV}{[Ne\,{\sc v}]\xspace}
\newcommand{\gleam}{\textsc{gleam}\xspace}
\newcommand{\lineergs}{10$^{40}$\,erg\,s$^{-1}$}
\newcommand{\sfrunit}{M$_{\odot}$\,yr$^{-1}$}
\newcommand{\kms}{km\,s$^{-1}$}

\received{August 12, 2022}
\revised{October 13, 2022}
\accepted{October 14, 2022}

\shorttitle{Host Galaxies of HyMoRS}
\shortauthors{Stroe et al.}

\begin{document}

\title{The Host Galaxies of Hybrid Morphology Radio Sources}

\correspondingauthor{Andra Stroe}
\email{andra.stroe@cfa.harvard.edu}

\author[0000-0001-8322-4162]{Andra Stroe}
\altaffiliation{Clay Fellow}
\affiliation{Center for Astrophysics \text{\textbar} Harvard \& Smithsonian, 60 Garden St., Cambridge, MA 02138, USA}

\author[0000-0002-4925-8403]{Victoria Catlett}
\affil{The University of Texas at Dallas, 800 W Campbell Rd, Richardson, TX 75080, USA}
\affiliation{Center for Astrophysics \text{\textbar} Harvard \& Smithsonian, 60 Garden St., Cambridge, MA 02138, USA}

\collaboration{}

\author[0000-0003-0251-6126]{Jeremy J. Harwood}
\affiliation{Centre for Astrophysics Research, School of Physics, Astronomy and Mathematics, University of Hertfordshire, College Lane, Hatfield, Hertfordshire AL10 9AB, UK}

\author[0000-0001-7093-3875]{Tessa Vernstrom}
\affiliation{ICRAR, The University of Western Australia, 35 Stirling Hw, 6009 Crawley, Australia}

\author[0000-0001-5649-938X]{Beatriz Mingo}
\affiliation{School of Physical Sciences, The Open University, Walton Hall, Milton Keynes, MK7 6AA, UK}

\begin{abstract}

Based on their differing radio morphologies, powerful radio galaxies can be separated into the Fanaroff-Riley I (FR-I) and II (FR-II) classes. Hybrid morphology radio sources (HyMoRS) contain morphologies consistent with each type of jet on either side: a powerful, highly relativistic FR-II-like jet terminating in a hotspot on one side and an FRI-like plume on the other. HyMoRS present a unique opportunity to study the conditions which give rise to the dichotomy. Using host galaxy properties, we conduct the first multiwavelength investigation into whether orientation can explain HyMoRS morphology. Through optical spectroscopy and mid-infrared photometry, we analyze the emission characteristics, and evaluate the broad characteristics of five HyMoRS host galaxies at intermediate redshifts ($0.4 < \mathrm{z} < 1.5$). The HyMoRS host galaxies in our sample have properties consistent with typical host galaxies of FR-II sources, suggesting that the observed hybrid morphologies may be caused by a dense, cluster-like environment bending FR-II jets combined with a favorable orientation which can make one side appear similar to an FR-I jet. Our results thus support the hypothesis that HyMoRS are mainly caused by environment and orientation.

\end{abstract}


\section{Introduction} 
\label{sec:intro}

The prominent radio emission in powerful active galactic nuclei (AGN) typically manifests through twin, symmetrical relativistic jets, spanning several hundreds of kpc to Mpc scales outward from their host galaxy. \citet{FR-1974} divided powerful radio sources into two morphological classifications, Fanaroff-Riley class I (FR-I) and class II (FR-II). In FR-I sources, the emission is brightest near the host galaxy, followed by extended diffuse emission outwards. On the other hand, FR-IIs feature prominent brightness peaks called ``hotspots" farther from the host galaxy with lobes of diffuse emission. The different morphology was later attributed to differing jet physics: unlike FR-I jets, which get disrupted quickly within the host and decelerate on tens of kpcs scales, FR-II jets stay collimated over large distances \citep{1994ApJ...422..542B, 2002MNRAS.336.1161L}.

\begin{table*}[ht!]
\centering
\caption{Positions, spectroscopic redshifts ($z_\mathrm{spec}$), spectrum origin instrument, SDSS magnitudes in the g, r, and i bands, and radio fluxes (at 1.4\,GHz and 4.9\,GHz) for each of the five HyMoRS in our sample.}
\label{tab:SDSS}
\begin{tabular}{ccccccccccc}
\toprule
Source & R.A. & Decl.  & $z_\mathrm{spec}$ & Instrument & g & r & i & $S_{1.4\,\mathrm{GHz}}$ & $S_{4.9\textrm{\,GHz}}$ \\
& $hh\,mm\,ss$ & $^{\circ}\,'\,''$ & & & mag & mag & mag & mJy &  mJy \\ 
\midrule
J1315+516 & 13\,14\,38.12 & 51\,34\,13.4 & 0.47799 & SDSS & $20.52\pm0.03$ & $19.75\pm0.02$ & $18.98\pm0.02$ & 93 & 51 	\\
J1348+286 & 13\,47\,51.58 & 28\,36\,29.6 & 0.74058 & BOSS &$17.39\pm0.01$ & $17.20\pm0.01$  & $17.43\pm0.01$ & 241 & 117 \\
J1313+507 & 13\,13\,25.78 & 50\,42\,06.2  & 0.88000 & GMOS & $21.45\pm0.05$ & $20.96\pm0.05$ & $20.36\pm0.04$ & 277 & 84 \\
J1154+513 & 11\,53\,46.43 & 51\,17\,04.1  & 1.37250 & GMOS & $22.10\pm0.12$  & $21.44\pm0.11$ & $21.37\pm0.16$ & 495 & 137 \\
J1206+503 & 12\,06\,22.39 & 50\,17\,44.3 & 1.45423 & BOSS & $21.38\pm0.04$ & $20.89\pm0.04$ & $20.69\pm0.05$ & 241 & 75 \\ 
\bottomrule
\end{tabular}
\end{table*}

The underlying cause of the FR-I/II dichotomy has been a topic of debate \citep[e.g.;][]{kaiser-1997}, and theories include both intrinsic and extrinsic causes. For example, the properties of the black hole engine could affect the magnetic fields which collimate and power the jets \citep[e.g.;][]{celotti-1997}, which could determine the initial jet velocities. The material in the host galaxy could affect the deceleration of the jet material \citep[e.g.;][]{2007MNRAS.381.1548K, 2019MNRAS.488.2701M}. This scenario is supported by the observed differences between the host galaxies of FR-I and FR-II sources \citep[e.g.;][]{2017MNRAS.466.4346M}. FR-I sources are hosted by massive red elliptical galaxies with spectra dominated by absorption features, no broad lines, and little-to-no narrow emission lines \citep[e.g.;][]{1964ApJ...140...35M, 2009ApJ...696..891H, 2012MNRAS.421.1569B, 2012A&A...541A..62J, 2018A&A...620A..16B}. At low jet power, both FR-Is and FR-IIs are predominantly low-excitation radio galaxies (LERGs) \citep{1994ASPC...54..201L, 2022MNRAS.511.3250M}. High-excitation radio galaxies (HERG) appear at high jet powers, in which FR-IIs dominate. HERGs, hosted by bluer galaxies with lower masses, higher star formation (SF) rates (SFR), and diskier morphologies, have strong narrow (an order of magnitude stronger than LERGs) and, in some cases broad, emission lines \citep[e.g.;][]{baum-1989, 1995ApJ...448..521Z, 2012MNRAS.421.1569B, 2017MNRAS.466.4346M, 2018A&A...620A..16B}. Finally, FR-Is tend to live in denser cluster environments than FR-II galaxies, suggesting that the density of the intergalactic/intracluster medium could decelerate the jets outside of the galaxy \citep[e.g.;][]{prestage-1988, 2019MNRAS.488.2701M}.

Hybrid morphology radio sources (HyMoRS) possess both FR-I and FR-II structures, presenting a unique opportunity to probe the conditions which cause the dichotomy. First identified in \citet{gopal-2000}, their existence suggests that the black hole engine cannot alone cause the morphological differences between FR-I's and FR-II's. The number of HyMoRS candidates has rapidly grown over the past few years, amounting to hundreds of sources and $\sim5\%$ of the resolved radio AGN population in modern radio surveys \citep{2019MNRAS.488.2701M, 2017AJ....154..253K}.

\begin{figure*}[ht!]
\centering
    \includegraphics[width=0.9\textwidth]{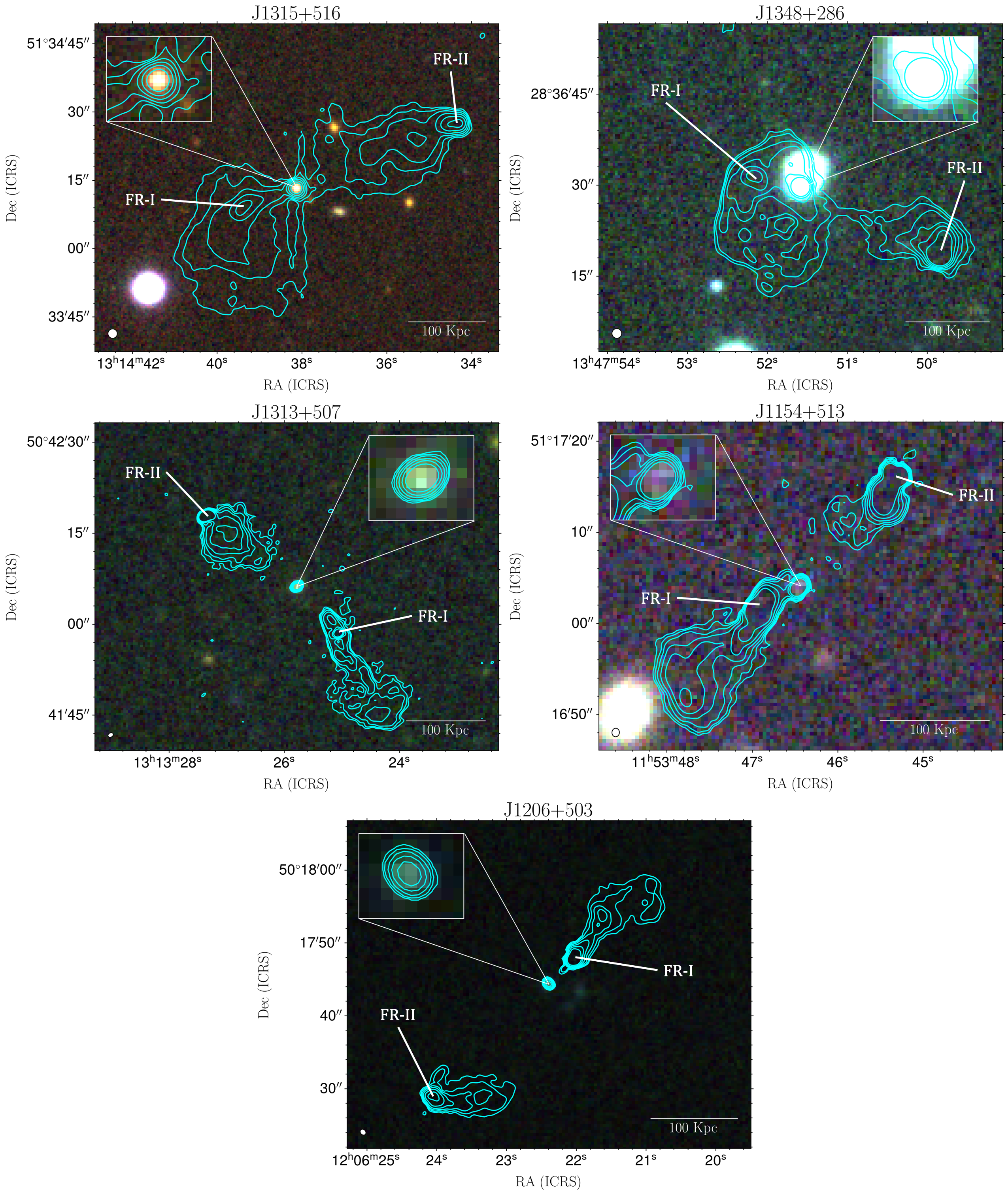}
    \caption{The five HyMoRS in our sample. The background images are Pan-STARRS RGB composites (using $g$, $r$, $i$ filters), with overlaid L-band VLA contours at $\sigma_{rms} \times 2^{n}$. Each image shows the host galaxy (zoomed-in section) with an FR-I-like plume on one side and an FR-II jet, hotspot, and lobe on the other.}
    \label{fig:radio-optical}
\end{figure*}

In this paper, using the intrinsic properties of the host galaxy, we investigate whether orientation can give rise to these morphological differences. As a relativistic jet from the black hole engine travels through the host galaxy, it may be slowed down by intervening material in a host galaxy or travel unimpeded and slow down on larger scales, eventually terminating in a bright hotspot \citep[e.g.;][]{2007MNRAS.381.1548K, 2019MNRAS.488.2701M}. The distribution and type of material in the galaxy could affect how quickly the jet slows, affecting the location of the final emission peak \citep{gopal-1996}. We aim to unveil the cause of the dichotomy by focusing our analysis on the best-studied five sources that \citet{gawronski-2006} securely identified as HyMoRS given their spatially-resolved radio observations (see Figure~\ref{fig:radio-optical}). L-band ($1.0-2.0$ GHz) images from the Very Large Array (VLA) show the presence of one FR-I-like and one FR-II set of structures in each system, but, using spectral aging techniques, \citet{harwood-2020} largely attribute the FR-I-like morphology to a favorable projection of an FR-II jet, hotspot, and lobe. We measure detailed galaxy properties by investigating their spectral properties with newly-obtained 1D spectroscopy from the Gemini Multi-Object Spectrograph North (GMOS-N) in combination with MIR color diagnostics. 

In Section \ref{sec:obs-red}, we present our new data taken with Gemini, as well as archival images and spectra, and we walk through the data reduction methods. Section \ref{sec:analysis} discusses how we analyze the spectral and ancillary data. In Section \ref{sec:results_discussion}, we discuss the insights into HyMoRS host galaxy properties and the implications for the broader formation context of powerful radio galaxies. Conclusions can be found in Section~\ref{sec:conclusion}.

We use the $\Lambda$CDM cosmological model of a flat universe with H$_{0}=\,$71\,km s$^{-1}$\,Mpc$^{-1}$, $\Omega_{m}=\,$0.27, and $\Omega_{\Delta}=\,$0.73. A \citet{1955ApJ...121..161S} initial mass function is used throughout.

\begin{figure*}
\centering
\label{bigfig:spectra}
{\includegraphics[width=0.95\textwidth]{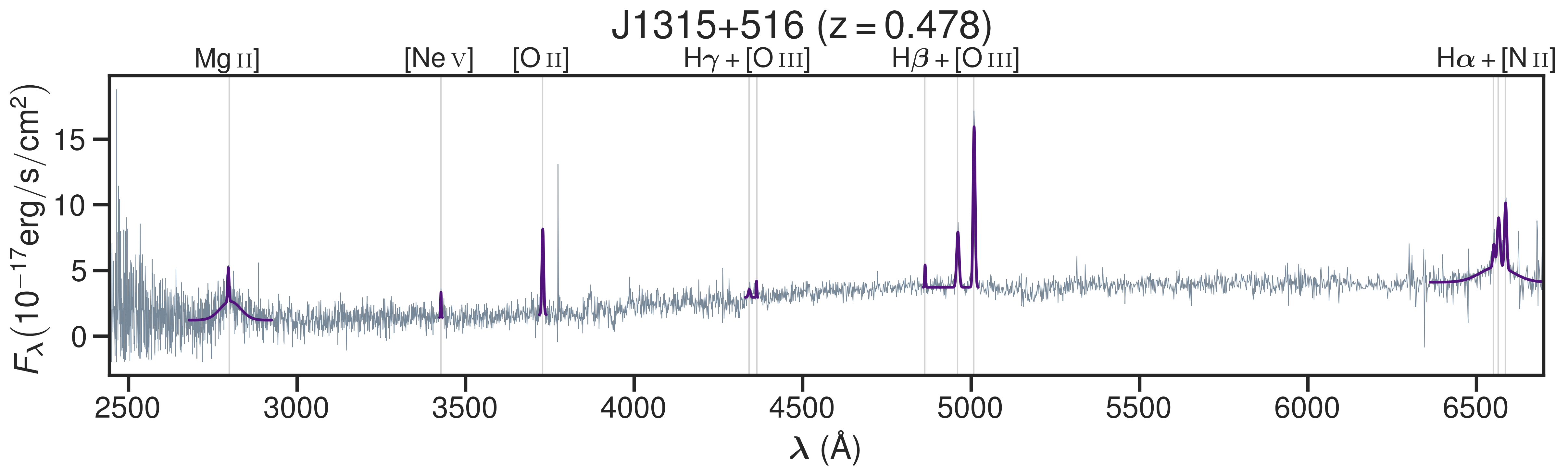}\\}
{\includegraphics[width=0.95\textwidth]{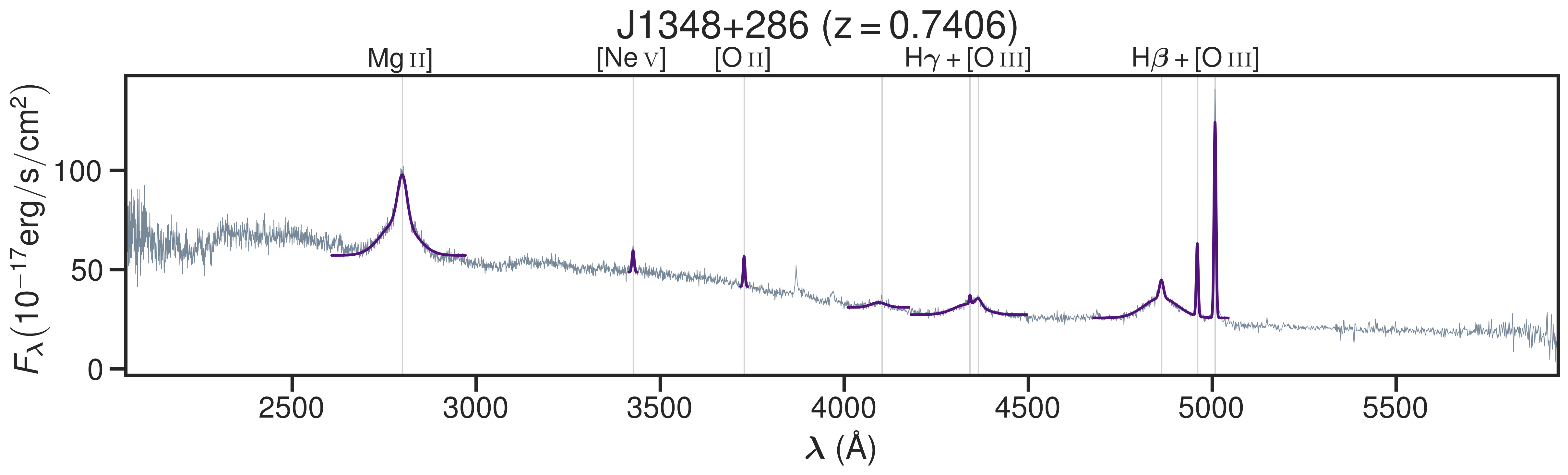}\\}
{\includegraphics[width=0.95\textwidth]{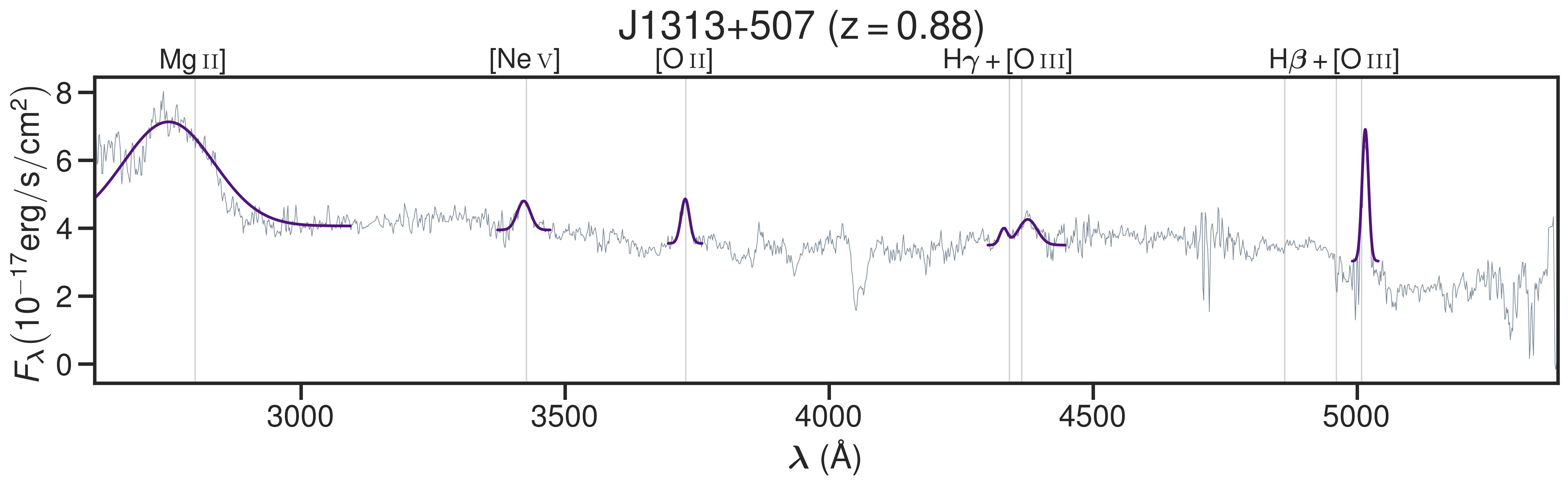}}\\
\caption{HyMoRS spectra with a selection of emission lines marked (gray vertical lines). Gaussian emission line fits are overplotted in the purple lines. All of our sources have prominent narrow and broad emission lines typical of Type 2 AGN and quasars, with \OII emission suggesting significant SFRs. The line luminosities and velocity FWHMs are listed in Tables \ref{tab:lines} and \ref{tab:fwhm}, respectively. }
\end{figure*}

\begin{figure*}\ContinuedFloat
\centering
{\includegraphics[width=0.95\textwidth]{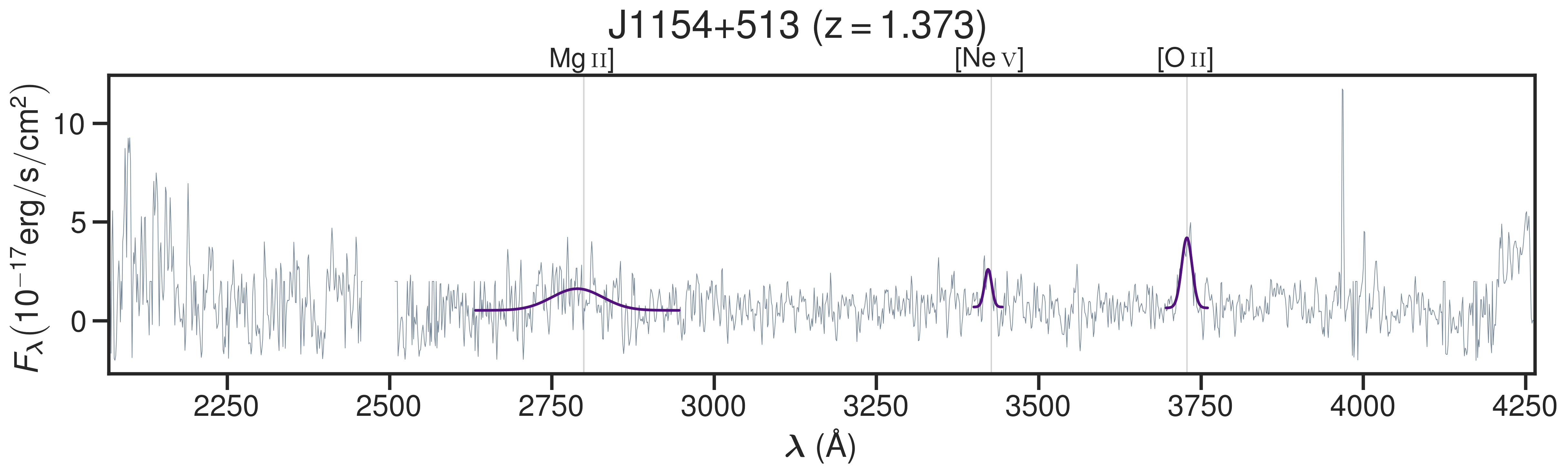}\\}
{\includegraphics[width=0.95\textwidth]{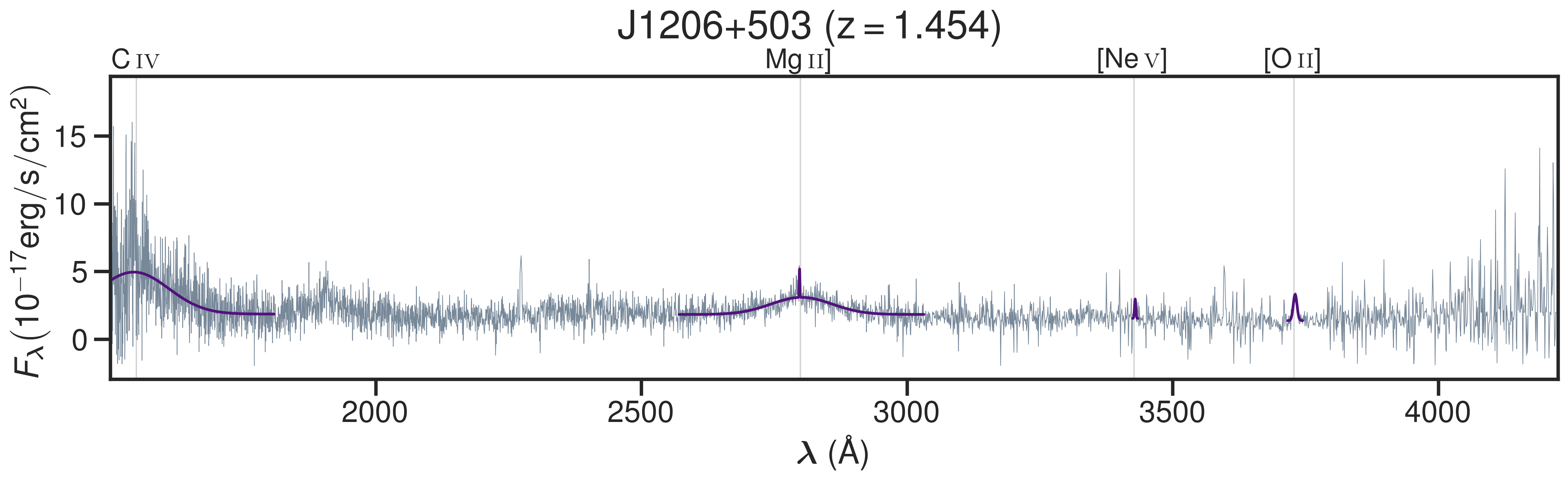}\\}
\caption{Continued.}
\end{figure*}

\section{Sample, Observations and Data Reduction} \label{sec:obs-red}

\subsection{HyMoRS sample}

The five sources in this study have been identified as bona-fide HyMoRS through a uniform radio selection by \citet{gawronski-2006}. After selecting all sources with 1.4\,GHz fluxes $>20$\,mJy and angular size $\theta>8^{''}$ in five random high galactic latitude $\approx16^\circ\times16^\circ$ areas within the VLA Faint Images of the Radio Sky at Twenty-centimeters (FIRST) survey \citep{1997ApJ...475..479W}, \citet{gawronski-2006} visually selected HyMoRS candidates and followed them with pointed, narrow-band VLA observations, resulting in a secure sample of five HyMoRS sources. The resolved spatial and spectral radio properties of these five sources were then studied in detail by \citet{harwood-2020} through high-resolution, wide-band (C, L), JVLA observations. In the present paper, we follow up these five best-studies sources, without imposing any additional selection criteria.

\subsection{GMOS-N Observations and Data Reduction}

The optical spectra from GMOS-N long-slit instrument are new to this paper (PI: Stroe). We observed the three sources without existing spectroscopic observations, J1313+507, J1154+513, and J1206+503, with GMOS-N in long-slit mode. To obtain a continuous wavelength coverage between 5,000\AA{} and 10,000\AA{}, we used the R150 grating centered at 8000 and 8500\,\AA{}, in combination with the GG455 filter, a $1^{\prime\prime}$ slit and $2\times2$ binning, resulting in 23\,\AA{} resolution. The data were collected in March 2019, taking advantage of the poorer observing conditions on Mauna Kea, with $\geq\,70-100$ percentile cloud coverage, and $\geq\,70-\geq\,85$ percentile overall image quality. J1313+507 and J1206+503 were observed over four exposures, for a total time on target of 44\,min. J1154+513 was observed for 55\,min, to compensate for the very poor cloud coverage during the last exposure. We also observed the standard star Hiltner 600 for flux calibration. Note that, subsequently, a higher quality BOSS spectrum for J1206+503 was made public (described below), which was used in the rest of the paper. 

We reduced the GMOS-N spectra from the Gemini North telescope for J1313+507 and J1154+513 according to the GMOS Data Reduction Cookbook using the Gemini IRAF package \citep{shaw-2016}. First, all exposures were debiased, and we flat-fielded the science frames and the standard star exposure. Next, we derived a wavelength solution for each instrumental setup by manually identifying emission lines in the arc spectra, resulting in a typical RMS of 5-8\,\AA{}. Next, we removed cosmic rays from the science exposures using Laplacian Cosmic Ray Identification (LACOS) from \citet{van-2001}. We then subtracted the sky in each exposure by using emission-free sky areas adjacent to the target trace. We extracted a 1D spectrum from each exposure using a 14$''$ area and median combined all four exposures for each target to improve the signal-to-noise ratio (S/N). Finally, we applied the sensitivity solution derived from an exposure of the standard star Hiltner 600, creating the final GMOS-N spectrum.

\subsection{Ancillary Data}

Spectroscopy for J1206+503, J1315+516, J1348+286 was extracted from the SDSS DR16 archives \citep{sdss} and used either the SDSS or the Baryon Oscillation Spectroscopic Survey (BOSS) spectrograph \citep{boss}. The SDSS/BOSS data cover the $\sim4,000-9,000$\,\AA{} range, at a resolution of 3.6\,\AA{}. 

To supplement the optical spectroscopy, we compiled optical and mid-IR photometry for all the sources using the VizieR Catalogue \citep{vizier}, including data from the Panoramic Survey Telescope and Rapid Response System \citep[Pan-STARRS][]{pan-starrs}, and the Wide-field Infrared Survey Explorer \citep[\textit{WISE}][]{wise,10.26131/IRSA1}.

\section{Analysis} \label{sec:analysis}
 
We created RGB composite images with radio contours using $g$, $r$, $i$ Pan-STARRS \citep{pan-starrs} and L-band VLA images \citep{harwood-2020}. The optical data enabled a clear identification of the host galaxy, while the FR-I -like jet and plume and the FR-II jet, hotspot and lobe were easily distinguished in the radio images.

\begin{table*}
\begin{center}
\caption{Luminosities for select UV and optical emission lines. Adjacent lines were fit together, and both narrow and broad components were fit where necessary.} \label{tab:lines}
\begin{tabular}{lcccccc}
\toprule
Line & $\lambda_\mathrm{vac}$ & J1315+516 & J1348+286 & J1313+507 & J1154+513 & J1206+503  \\
& \AA{} & \lineergs & \lineergs & \lineergs & \lineergs & \lineergs \\
\midrule
\CIV & 1549.48 &  &  &  &  & $6876\pm1326$ \\
\CIII (narrow) & 1908.734 &  &  &  &  & $70\pm23$ \\
\CIII (broad) & 1908.734 &  &  &  &  & $1276\pm148$ \\
\CII (narrow) & 2326.0 &  & $<25$ &  & $<309$ & $<19$  \\
\CII (broad) & 2326.0 &  & $821\pm138$ &  & $<309$ & $238\pm72$ \\
\MgII (narrow) & 2799.12 & $12\pm3$ & $1816\pm146$ & $<17$ & $<83$ & $44\pm21$ \\
\MgII (broad) & 2799.12 & $93\pm12$ & $5032\pm183$ & $2533\pm108$ & $1252\pm279$ & $2445\pm219$ \\
\NeV & 3426.85 & $4\pm2$ & $198\pm24$ & $102\pm15$ & $305\pm54$ & $60\pm21$ \\
\OII & 3728.483 & $36\pm2$ & $234\pm36$ & $100\pm10$ & $827\pm61$ & $228\pm29$ \\
\NeIII & 3868.76 & $10\pm3$ & $261\pm44$ & $26\pm9$ & $<104$ & $<43$  \\
\Hdelta & 4102.89 & $<2$ & $316\pm80$ & $<14$ &  & $<102$  \\
\Hgamma (narrow) & 4341.68 & $4\pm2$ & $62\pm17$ & $33\pm15$ &  &  \\
\Hgamma (broad) & 4341.68 & $<1$ & $1321\pm94$ & $<14$ &  &  \\
\Hb (narrow) & 4862.68 & $5\pm1$ & $310\pm36$ & $<23$ &  &  \\
\Hb (broad) & 4862.68 & $<1$ & $2884\pm100$ & $<23$ &  &  \\
\OIII & 4960.295 & $33\pm2$ & $610\pm73$ & $<24$ &  & \\
\OIII & 5008.24 & $81\pm3$ & $1729\pm100$ & $229\pm21$ & &  \\
\NII & 6549.86 & $11\pm4$ &  &  &  & \\
\Ha (narrow) & 6564.614 & $33\pm5$ &  &  &  & \\
\Ha (broad) & 6564.614 & $124\pm18$ &  &  &  & \\
\NII & 6585.27 & $35\pm4$ &  &  &  & \\
\bottomrule
\end{tabular}
\end{center}
\end{table*}

\subsection{Spectral Lines}

To probe SF and AGN activity within the sources, we identified prominent rest-frame optical and UV emission lines in each source using the Galaxy Line Emission \& Absorption Modeling (\gleam) software \citep{gleam}. For each potential emission line, \gleam searches for emission peaks near the expected central wavelength, omitting two ranges that feature prominent telluric absorption in our analysis ($7586-7658$\,\AA{} $6864-6945$\,\AA{}). Once it identifies an emission line and its adjacent continuum, \gleam fits a Gaussian profile to the emission to derive parameters, such as the total flux and equivalent width. Lines located nearby are jointly fit, and \CIII, \CII, \MgII, \Hgamma, \Hb, and \Ha were fit with a two-component Gaussian, as required to identify any broad and narrow components. A line was considered a detection when S/N$>3$. For undetected lines, we calculated 3$\sigma$ upper flux limits. The luminosities of the fitted lined are listed in Table \ref{tab:lines}. For resolved lines, we also calculate the velocity FWHM $v_\mathrm{FWMH}$ (see Table~\ref{tab:fwhm}).

\begin{figure}[htb!]
    \centering
    \includegraphics[width=0.5\textwidth]{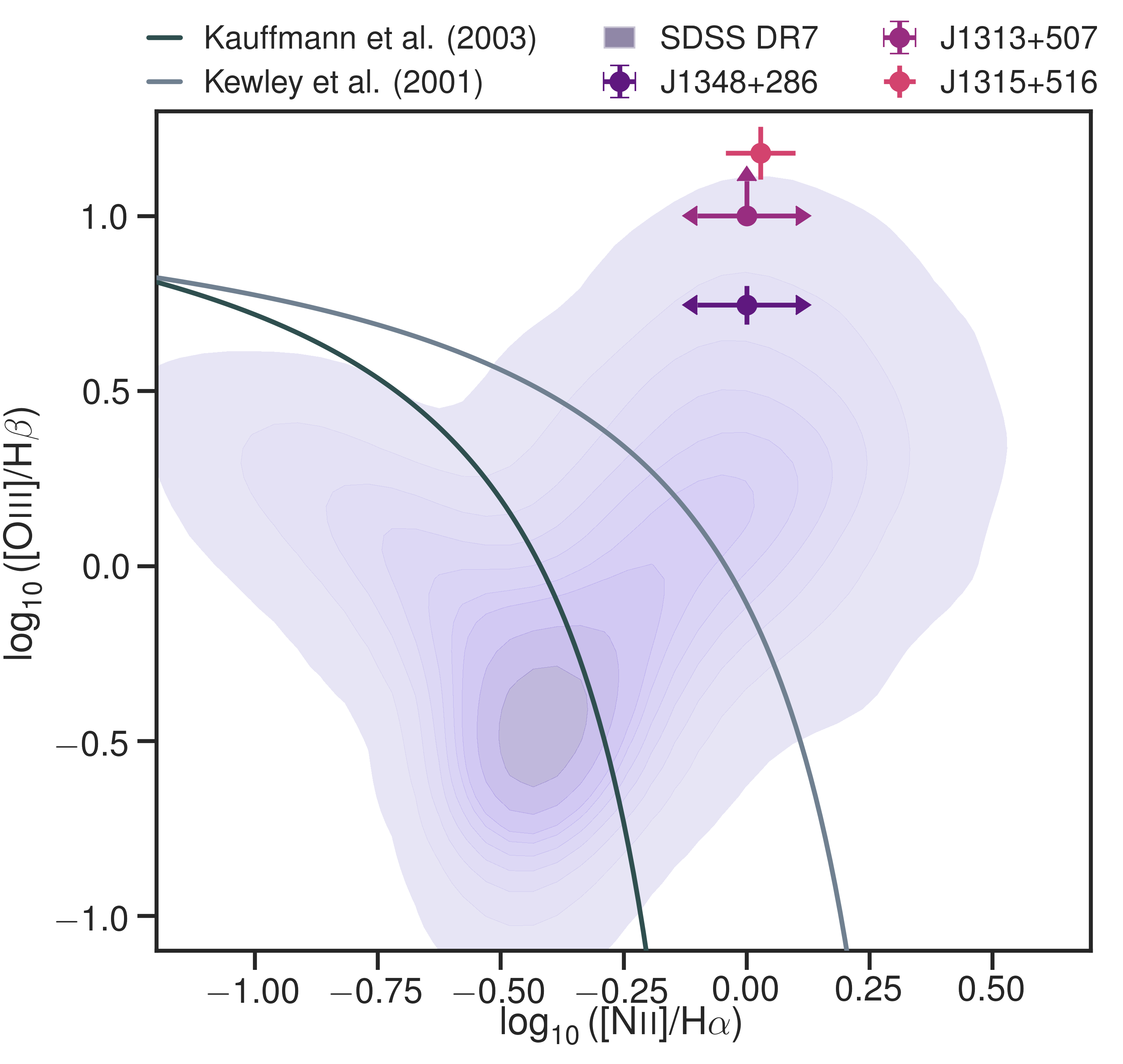}
    \caption{The \citet{kauffmann-2003} and \citet{kewley-2001} lines separate star-forming galaxies, composites and AGN in the BPT diagram \citep{baldwin-1981}. The high \OIII/Hb narrow-line ratios securely classify J1348+286, J1313+507, and J1315+516 as Seyferts, even in the absence of \NII/\Ha coverage. Given their high redshift of $z>1$, none of the four required emission lines are covered for J1206+503 and J1154+513. The contours represent 10,000 sources sampled from SDSS DR7 \citep{abazajian-2009}.}
\label{fig:bpt}
\end{figure}

As well as the velocity width of the lines, the narrow line components of the \NII/\Ha and \OIII/\Hb ratios can be used to distinguish between star-forming galaxies and AGN \citep[see Figure~\ref{fig:bpt};][BPT diagram]{baldwin-1981}. We included sample galaxy data from SDSS DR7 on this diagram for comparison \citep{abazajian-2009}. 

\begin{table*}
\begin{center}
\caption{Deconvolved emission line velocity widths.} \label{tab:fwhm}
\begin{tabular}{lcccccc}
\toprule
Line & $\lambda_\mathrm{vac}$ & J1315+516 & J1348+286 & J1313+507 & J1154+513 & J1206+503\\
 & $\mathrm{\mathring{A}}$ & $\mathrm{km\,s^{-1}}$ & $\mathrm{km\,s^{-1}}$ & $\mathrm{km\,s^{-1}}$ & $\mathrm{km\,s^{-1}}$ \\
 \midrule
\CIV & 1549.48 &  &  &  &  & $30066\pm4698$ \\
\CIII (narrow) & 1908.734 &  &  &  &  & $336\pm138$ \\
\CIII (broad) & 1908.734 &  &  &  &  & $12805\pm1469$ \\
\CII & 2326.0 &  & $8648\pm1429$  &  &  & $4358\pm1401$ \\
\MgII (narrow) & 2799.12 & $460\pm140$ & $3126\pm174$ & & & \\
\MgII (broad) & 2799.12 & $7736\pm1003$ & $11457\pm445$ & $21664\pm748$ & $9836\pm2051$ & $14533\pm1051$ \\
\NeV & 3426.85 & $95\pm59$ & $579\pm81$ & $2332\pm393$ & $754\pm225$ & $230\pm87$ \\
\OII & 3728.483 & $455\pm27$ & $456\pm81$ & $1148\pm166$ & $1299\pm124$ & $654\pm99$ \\
\NeIII & 3868.76 & $525\pm123$ & $656\pm128$ &  &  & \\
\Hdelta & 4102.89 &  & $3569\pm936$ & $9589\pm2102$ & \\
\Hgamma (narrow) & 4341.68 & $448\pm257$ & $344\pm98$ & $860\pm555$ & & \\
\Hgamma (broad) & 4341.68 &  & $6403\pm375$ &  & \\
\OIII & 4364.436 &  & $1402.0\pm288.0$ & $2730\pm573$ & & \\
\Hb (narrow) & 4862.68 & $158\pm42$ & $795\pm83$ &  & \\
\Hb (broad) & 4862.68 &  & $6641\pm241$ &  &  \\
\OIII & 4960.295 & $508\pm35$ & $362\pm49$ & \\
\OIII & 5008.24 & $413\pm13$ & $378\pm22$ & $468\pm88$ & \\
\Ha (narrow) & 6564.614 & $430\pm61$ &  &  & & \\
\Ha (broad) & 6564.614 & $5347\pm773$ &  &  &  & \\
\NII & 6585.27 & $337\pm37$ &  &  &   & \\
\bottomrule
\end{tabular}
\end{center}
\end{table*}

\subsection{Star formation rates}

\Ha cannot reliably be used to estimate the SFR of the host galaxy when an AGN is present. However, we can use \OII luminosities to roughly estimate SFRs, since \OII is not excited in the broad-line region and only weakly excited in the narrow-line region \citep{2001AJ....122..549V}. However, AGN can induce \OII in extended emission line regions (EELRs) throughout the galaxy, which can bias high our SFRs \citep{2018MNRAS.480.5203M}.

We used the \OII luminosity to SFR the conversion from \citet{kennicutt-1998}:

\begin{equation}
    SFR = (1.4 \pm 0.4) \times 10^{-41} L_{[\mathrm{O}\,\textsc{ii}]}
\end{equation}

\noindent where SFR is measured in M$_{\odot}$\,yr$^{-1}$ and L$_{\mathrm{O}\textsc{ii}}$ is the total luminosity of the \OII emission line in erg\,s$^{-1}$. 

We compared the SFRs to values for typical star-forming galaxies ($SFR^*$) at the same redshift using the parametrization from \citet{2014MNRAS.437.3516S}:
\begin{equation}
    SFR^*(z)=10^{0.55*z+0.57}
\end{equation}
where $SFR^*$ is measured in M$_{\odot}$\,yr$^{-1}$ and $z$ is the redshift.

The SFRs of the sources and the typical SFRs at their redshift are listed in Table \ref{tab:sfr}. Using \NeV to measure the EELR contribution in a stack of radio-loud quasars, \citet{2018MNRAS.480.5203M} found that $\approx50$\,\% of the \OII emission can be attributed to SF and $\approx50$\,\% to EELRs. Three of our sources (J1315+516, J1154+513, J1206+503) have \OII/\NeV ratios higher than the averages from the \citet{2018MNRAS.480.5203M}, ranging from $2.7-9$, indicating a lower contamination from EELRs. On the other hand, our \OII luminosities might be underestimated because no dust extinction correction was applied.

\begin{table}
\begin{center}
\caption{\OII-derived SFRs of the HyMoRS. No corrections for contamination from EELRs and dust absorption were applied. For comparison, we list the typical SFRs of star-forming galaxies at the redshift of each source.} \label{tab:sfr}
\begin{tabular}{lcc}
\toprule
Source & SFR & $SFR^*(z)$\\
 & $\mathrm{M_{\odot}\,yr^{-1}}$ & $\mathrm{M_{\odot}\,yr^{-1}}$ \\
 \midrule
J1315+516 & $5.1\pm1.5$ & 6.8\\
J1348+286 & $33.8\pm10.9$ &  9.5\\
J1313+507 & $14.0\pm4.2$ &  11.3\\
J1154+513 & $115.8\pm34.2$ & 21.1\\
J1206+503 & $31.9\pm 9.9$ &  23.4\\
\bottomrule
\end{tabular}
\end{center}
\end{table}

\subsection{Mid IR color-color plots}

\begin{figure}[htb!]
    \centering
    \includegraphics[width=0.5\textwidth]{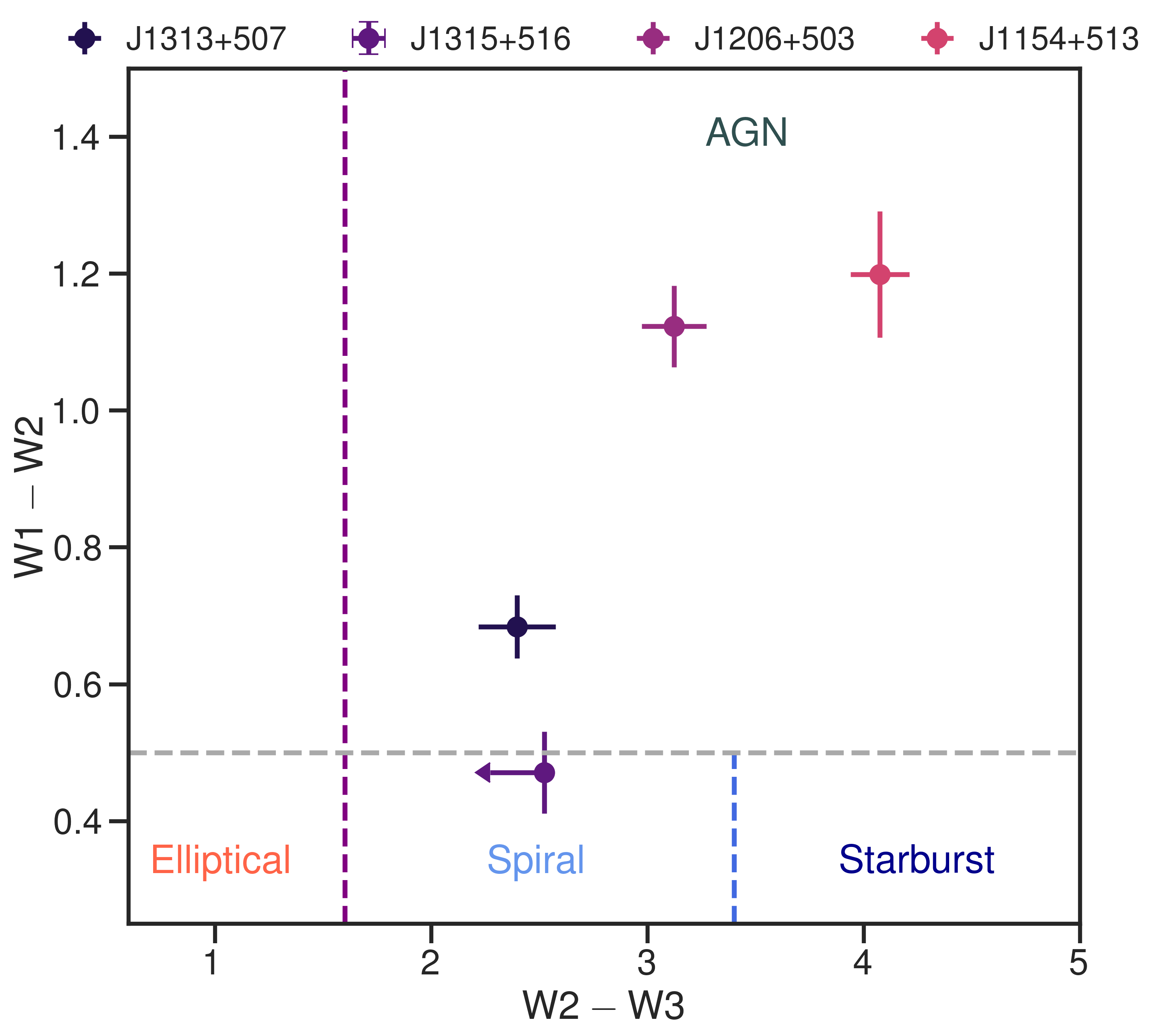}
    \caption{\textit{WISE} W1-W2 versus W2-W3 color-color plot, including the \citet{2016MNRAS.462.2631M, 2022MNRAS.511.3250M} lines which separate elliptical, spiral, and starburst galaxies, and AGN. The four sources with secure \textit{WISE} measurements are securely located in the AGN quadrant of this plot, where the bulk of FR-IIs reside \citep{2022MNRAS.511.3250M}.}
\label{fig:midIR}
\end{figure}

We employ \textit{WISE} $\mathrm{W2}-\mathrm{W3}$ versus $\mathrm{W1}-\mathrm{W2}$ color-color plots to separate AGN from galaxies, using the classification lines from \citep{2016MNRAS.462.2631M}. We used profile-fit photometry magnitudes without aperture corrections, given the high redshifts of our sources. J1348+286 could not be included in this analysis because its proximity to another source impeded the proper deblending in the WISE data. The \textit{WISE} color-color plot can be found in Figure~\ref{fig:midIR}

\section{Results and Discussion}
\label{sec:results_discussion}

\subsection{Nature of the HyMoRS host galaxies}

In this section, we combine insights from the color-color properties and the spectral analysis to obtain a consistent picture of the nature of our five sources. For detailed notes on each source, please see Section~\ref{sec:notes}.

The optical spectroscopy brings additional evidence to hone in on the AGN nature of the sources. Every source presents narrow lines and at least two broad ($v_\mathrm{FWHM}>1000$\,\kms) emission lines, as shown in Table \ref{tab:fwhm}, securely identifying all sources as Type 1 AGN or quasars. Using the BPT diagram, we classify the narrow-line emission for the three lower redshift sources as dominated by the AGN narrow line region instead of SF from \HII regions (Figure~\ref{fig:bpt}). The Lorentzian, blue-shifted, and/or winged \OIII profiles indicate significant turbulence and fast-moving clouds in the narrow line regions and outflows. Moreover, all the sources have detections of \NeV, which, with its very high 97\,eV ionization potential, implies a hard ionizing spectrum decidedly associated with nuclear activity \citep{2016MNRAS.456.3354F}. In our highest redshift source, the asymmetric \CIV could be caused by non-gravitational effects, such as dust scattering and radiation pressure as proposed by \citet{2005MNRAS.356.1029B}. Among the five sources, J1348+286 has the strongest evidence for a quasar classification, given its power-law continuum. All four sources with \textit{WISE} measurements are securely classified as AGN in the color-color diagnostic plot (see Figure~\ref{fig:midIR}). The sources occupy the same part of the color space ($(\mathrm{W1}-\mathrm{W2})>0.5$) as the bulk of HERGs and high-luminosity FR-II sources (i.e. FR-IIs located above the canonical $L_{150\,\mathrm{MHz}} \sim 10^{26}$\,W\,hz$^{-1}$ line) \citep{2022MNRAS.511.3250M}. By contrast, FR-Is and lower luminosity FR-IIs are almost exclusively located below the 0.5 line \citep{2022MNRAS.511.3250M}. Our sources also fall within the AGN wedge of \citep{2012MNRAS.426.3271M}. Despite being clear AGN, our HyMoRS host galaxies display significant \OII SFRs, typical of star-forming galaxies at their redshift.  Therefore, even assuming 50\% contamination of the \OII fluxes by EELRs, our sources still have significant SFRs, which are comparable or higher than typical $SFR^*$ at their redshift ($SFR/SFR^*\approx0.4-2.7$).

\subsection{Are HyMoRS true FR-I/II hybrids?}

Among the various HyMoRS formation models, we aim to determine whether the properties of the host galaxy could cause each of its radio jets to take on a different FR morphology. High excitation emission lines are found almost exclusively in the spectra of high radio luminosity FR-IIs \citep{2022MNRAS.511.3250M, 2017A&ARv..25....2P}. In line with our findings, these sources also have significant SFRs, indicative of ample gas supplies \citep{2022MNRAS.511.3250M}. Given their SFRs ranging from $2.5-58$\,\sfrunit, our HyMoRS are hosted by typical star forming galaxies. Supporting the FR-II classification based on the high radio luminosities ($L_{\rm178\,MHz}\sim10^{27}$\,W\,Hz$^{-1}$\,sr$^{-1}$ which is well above the $10^{25.5}$\,W\,Hz$^{-1}$\,sr$^{-1}$ FRI/FRII division line) and the spatially-resolved spectral properties \citep{harwood-2020}, the optical and midIR data from the present paper paint a consistent picture: our HyMoRS are HERGs whose hosts are star-forming galaxies with strong Type 1 AGN / quasar features. Since radio quasars and Type 1 AGN are intrinsically radio galaxies with a narrower viewing angle \citep{2017A&ARv..25....2P}, our results thus indicate that, in a simplified scenario, HyMoRS might effectively be FR-IIs in which the small angle between the radio jet and the line of sight of $<45^\circ$ enables a clear view of both the narrow and broad line regions. In reality, while the majority of Type 1 AGN are relatively unobscured, the torus covering factor can vary between individual AGN, resulting in a significant overlap between the covering factor distributions between Type 1 and Type 2 AGN \citep[e.g.][]{2012ApJ...747L..33E, 2019ApJ...872..168S}.

When originally discovered \citep{gawronski-2006}, and in subsequent large-scale survey searches \citep[e.g.,][]{2017AJ....154..253K, 2019MNRAS.488.2701M}, HyMoRS candidates presented a compelling case for the existence of a hybrid FR-I/II source. However, these initial searches didn't benefit from wide-band, multi-frequency data needed to disentangle their detailed morphology, especially if significant projection effects are involved. \citet{harwood-2020} demonstrated that, in order to unequivocally classify the two sides as an FR-I or FR-II, detailed radio spectral analyses are required. \citet{harwood-2020} posited that the hybrid morphologies are actually caused by a combination of large-scale environmental effects pushing back FR-II jets and a favorable orientation \citep[see Figure 4 in][]{harwood-2020}. In another detailed multiwavelength study of a giant FR-II radio galaxy which presents some HyMoRS-like asymmetry between the lobes, \citet{2020PASA...37...13S} conclude that the FR-I-like jet is propagating into the dense intra-cluster medium (ICM) of the nearby irregular cluster, which slows it down. Such processes would compound with the ram pressure the ICM can exert on the radio jets, resulting in unusual morphologies that can be interpreted as hybrid sources when seen in projection. 
Our results seem to align with the conclusions from the detailed radio studies from \citet{harwood-2020} and \citet{2020PASA...37...13S}. Unlike typical powerful FR-IIs, which tend to be found, on average, in lower density environments than FR-Is \citep[e.g.][]{2017A&ARv..25....2P, 2019MNRAS.488.2701M}, the scenario that best fits the available data is that FR-II HyMoRS would more likely be found in high-density environments, such as a low-mass cluster or group. The bent nature of the radio jets, as suggested by the scenario proposed by \citet{harwood-2020} implies significant relative motions between the HyMoRS and the large-scale environment responsible for the bending. \citep{2017AJ....154..253K}, for example, find that one of their sample of 25 HyMoRS is hosted by a cluster. Given the high SFRs, which indicate the presence of gas, the host galaxies could not have experienced much quenching as is typical for an overdense environment, such as a cluster. However, if HyMoRS are recent infallers into a cluster, that could explain both the relative motion and the star-forming nature of the host galaxy. To further test this hypothesis, we searched for evidence of overdensities within 30\,arcmin of the HyMoRS. The 10\,ks \textit{Chandra} exposure (PI: Kraft) targeting J1315+516 unveils a low surface brightness, extended area ($\sim5\times5$\,arcmin$^2$) towards the North-East of the source, with count density $\sim1.5$ times about the rest of the field. While this could indicate the presence of a group or cluster, a much deeper observation would be needed to conclusively determine whether an ICM is present. J1348+286 has accidental X-ray coverage with \textit{Chandra} and \textit{XMM-Newton}, but the low exposures times ($<16$\,ks) prevent a clear test of the overdensity hypothesis. Sunyaev-Zel'dovich detections in all-sky surveys, with, e.g., Planck, and optical search algorithms in all-sky photometric survey data have increasing errors and lower detection rates with redshift, high cluster mass limits \citep[$\sim4\times10^{14}$\,M$_{\odot}$ at $z>0.5$][]{2016A&A...594A..27P}, as well as redshift upper limits for cluster detections \citep[0.55 for redMaPPer][]{2014ApJ...785..104R}, which prevent a clear conclusion regarding the presence of a cluster or group around our high redshift HyMoRS. The quality of the available data limits our ability to draw clear insights into the presence or absence of an overdensity around the HyMoRS, with deep pointed observations being required. 

\section{Conclusions}\label{sec:conclusion}

In this work, we presented the first investigation into the nature of HyMoRS host galaxies with the goal of shedding light on the formation processes that drive the FR-I/II dichotomy. Using primarily optical spectroscopy, we studied the properties of the host galaxies of the five best-studied HyMoRS, which benefit from detailed radio investigations into their hybrid nature. We conclude that the HyMoRS in  our sample are hosted by star-forming, disky galaxies with HERG spectral features typical of Type 1 AGN and quasars which are almost exclusively found in powerful FR-II sources. This sample of five was chosen as the most well defined and best studied HyMoRS examples. Though subsequent studies have identified more candidates using varied criteria, a larger follow-up study would be necessary to confirm that the conclusions here hold for other the HyMoRS population as a whole. Future deep, pointed, X-ray observations could further test the scenario that HyMoRS are infalling into a cluster that bends the jets because of the relative motion to the ICM. As proposed by \citet{harwood-2020}, the hybrid morphology is likely caused by a favorable viewing angle which makes one of the FR-II jets appear to have an FR-I configuration, as well as enables a clear view of the narrow and broad line region resulting in a Type 1 AGN/quasar optical classification. We thus predict that, unlike the broader FR-II population, HyMoRS are likely to be found in galaxy cluster environments. However, deep, pointed X-ray observations are needed to fully test this hypothesis.

\citet{gopal-2000} theorized that the existence of HyMoRS indicates that the black hole engine cannot alone determine the location of its jet hotspot, the morphological difference between the FR-I and FR-II classes. However, our results, albeit limited to a small sample of five, indicate that we have yet to find bona fide FR-I/II hybrid sources. If all HyMoRS are bent FR-II sources, this re-opens the question about how the black hole engine influences the jets, further supporting findings from modern large-scale radio surveys \citep{2022MNRAS.511.3250M}. 

\section{Acknowledgments}\label{sec:acknowledgments}

We thank the referee for their thoughtful comments which improved the paper. We also thank Matthew Ashby and Jonathan McDowell for stimulating discussions on potential avenues for the project. Andra Stroe gratefully acknowledges the support of a Clay Fellowship. Victoria Catlett also acknowledges the Smithsonian Astrophysical Observatory REU program, which is funded in part by the National Science Foundation REU and Department of Defense ASSURE programs under NSF Grants no.\ AST 1852268 and 2050813, and by the Smithsonian Institution. Beatriz Mingo acknowledges support from the Science and Technology Facilities Council (STFC) under grant ST/T000295/1. We recognize and acknowledge the very significant cultural role and reverence that the summit of Maunakea has always had within the indigenous Hawaiian community. We are most fortunate to have the opportunity to conduct observations from this mountain. This publication makes use of data products from the Wide-field Infrared Survey Explorer, which is a joint project of the University of California, Los Angeles, and the Jet Propulsion Laboratory/California Institute of Technology, funded by the National Aeronautics and Space Administration.

\vspace{5mm}
\facilities{Gemini North (spectroscopy), Pan-STARRS (survey imaging, photometry), Sloan (SDSS survey imaging, photometry, spectroscopy), VLA (imaging), WISE (photometry)}

\software{
  GLEAM \citep{gleam},
  Astropy \citep{astropy},
  APLpy \citep{aplpy},
  DS9 \citep{ds9},
  Matplotlib \citep{matplotlib},
  SciPy \citep{scipy}
}

\appendix

\section{Notes on individual sources}\label{sec:notes}

\subsection{J1315+516}
Both the narrow and broad emission lines confirm this source as AGN-dominated. The narrow line ratios firmly place the source in the Seyfert part of the BPT diagram (Figure~\ref{fig:bpt}). The AGN nature of the source is further supported by the broad \MgII and \Ha components, with velocity FWHM of over $5000$\,\kms (Table~\ref{tab:fwhm}). The source also has a significant detection of the very high ionization potential \NeV line. The source also has blue-shifted outflows as indicated by the winged \OIII$\lambda5008.24$ line. While the host galaxy is optically red, the galaxy has significant SF activity as traced by the \OII emission, in line with typical star-forming galaxies at its redshift \citep[cf.][]{2014MNRAS.437.3516S}. 

\subsection{J1348+286}

J1348+286 is the brightest source in our sample, with emission line luminosities $>10^{43}$\,erg\,s$^{-1}$. It has a point-like, quasar morphology in the optical imaging. The high S/N BOSS spectrum has a power-law continuum and strong \MgII and \Hb line detections, which contain both broad (including \MgII $v_{\mathrm{FWHM}}>11,000$\,\kms, \Hdelta at over 3,000\,\kms, and \Hgamma and \Hb at over 5,000\,\kms, Table~\ref{tab:fwhm}) and narrow components, consistent with an orientation in which both the narrow and the broad line regions are visible. Given its extreme \OIII/\Hb ratio and the likely \NII/\Ha ratio ranges, the source is securely placed in the Seyfert section of the BPT diagram (Figure~\ref{fig:bpt}). The \OIII$\lambda5008.24$ emission has a Lorentzian shape with wings on either side, indicating blue- and redshifted outflows.

\subsection{J1313+507}

A Type 2 source, J1313+507 has extremely broad, blue-shifted \MgII emission, with $v_{\mathrm{FWHM}>20,000}$\,km\,s$^{-1}$ (Figure~\ref{bigfig:spectra}, Table~\ref{tab:fwhm}). The presence of \NeV emission also confirms the AGN nature. The high $S/N\sim10$ \OIII detection and no \Hb detection enable us to place a lower limit on the \OIII/\Hb ratio indicating strong AGN contributions (Figure~\ref{fig:bpt}). The host galaxy SFR ($13.9\pm4.2$\,$\mathrm{M_{\odot}\,yr^{-1}}$) is consistent with typical star-forming galaxies at its redshift \citep[11.3\,$\mathrm{M_{\odot}\,yr^{-1}}$][]{2014MNRAS.437.3516S}.

\subsection{J1154+513}
J1206+503 is the highest redshift source in our sample and has the lowest S/N spectrum and faintest optical host ($g>22\,\mathrm{mag}$). Despite the low S/N, we detect broad \MgII and \NeV, which confirms that the source is an AGN (Figure~\ref{bigfig:spectra}, Table~\ref{tab:fwhm}). The \OII-derived SFR of $115.8\pm34.2$\,$\mathrm{M_{\odot}\,yr^{-1}}$ is over 5 times higher than $SFR^*$, classifying J1154+513 as a strongly star-forming source (Table~\ref{tab:sfr}). 

\subsection{J1206+503}
We detect \CIV and \MgII with clearly broad profiles characteristic of AGN emission ($>30,000$ and $>14,000$\,\kms, respectively). \CIV is asymmetric, with a blue excess, with \MgII containing a broad and narrow component.

\end{document}